\documentclass[a4paper,fleqn,usenatbib]{mnras}

\usepackage{newtxtext,newtxmath}

\usepackage[T1]{fontenc}
\usepackage{ae,aecompl}

\usepackage{graphicx}	
\usepackage{amsmath}	
\usepackage{amssymb}	

\title['Oumuamua's elongation by dust bombardment]{Dust bombardment can explain the extremely elongated shape of 1I/'Oumuamua and the lack of interstellar objects}

\author[D. E. Vavilov, Y. D. Medvedev]{
	Dmitrii E. Vavilov,\thanks{E-mail: vavilov@iaaras.ru}
	Yurii D. Medvedev
	\\
	The Institute of Applied Astronomy of the Russian Academy of Sciences, Kutuzova emb., St. Petersburg 191187, Russian Federation
}

\date{Accepted XXX. Received YYY; in original form ZZZ}

\pubyear{2018}

\begin{document}
	\label{firstpage}
	\pagerange{\pageref{firstpage}--\pageref{lastpage}}
	\maketitle
	
	\begin{abstract}
		Asteroid 1I/'Oumuamua is the first observed interstellar object. Its light-curve amplitude indicates that the object is highly elongated with an axis ratio of at least 5:1. The absence of such elongated asteroids in the Solar system, as well as the apparent lack of observed interstellar objects, are intriguing problems. Here we show that 'Oumuamua may have originated as a  slightly-elongated asteroid about $500\times300$ meters in size. 
		 {Surface} erosion, caused by interstellar dust bombardment, subsequently increased the axis ratio. Simply traveling through the interstellar medium for 0.03 to 2 Gyrs would have sufficed to give 1I its present shape. Passing through  a 10 pc dust cloud with a grain density of $10^{-23} \ \mathrm{g/cm^3}$ at 50 km/s would have had a similar effect on 'Oumuamua's form. Smaller objects of around 100 meters in diameter can travel the Galactic disk for merely 30 Myrs before they are disrupted. This could explain the small number of interstellar objects observed to date.
	\end{abstract}
	
	
	
	
	\section{Introduction}

	1I/'Oumuamua, the only confirmed interstellar object,~\citep{Meech_etal2017} was first observed by the Panoramic Survey Telescope and Rapid Response System 1 (Pan-STARRS 1) on 2017 October 19. Precovery data from  Pan-STARRS 1 and the subsequent follow-up campaigns allowed for a preliminary orbit determination identifying an object that was  leaving our Solar System. The object turned out to be quite unique having an orbital eccentricity of around 1.2\footnote{\label{foot:MPC}\url{https://www.minorplanetcenter.net/mpec/K17/K17UI1.html.}}. Such a high eccentricity indicates that the object was on an unbound hyperbolic orbit and most likely originated from outside our Solar System. Due to its highly eccentric orbit the Minor Planet Center classified 'Oumuamua firstly as a comet$^{\ref{foot:MPC}}$. The lack of cometary activity in images taken of the object by several observatories~\citep{Meech_etal2017,Fitzsimmons_etal2018,Ye_etal2017} has seen the object's designation altered to ''1I/'Oumuamua'' , the ''messenger'' from the stars.
	
	\begin{figure}
		\includegraphics[width=\columnwidth]{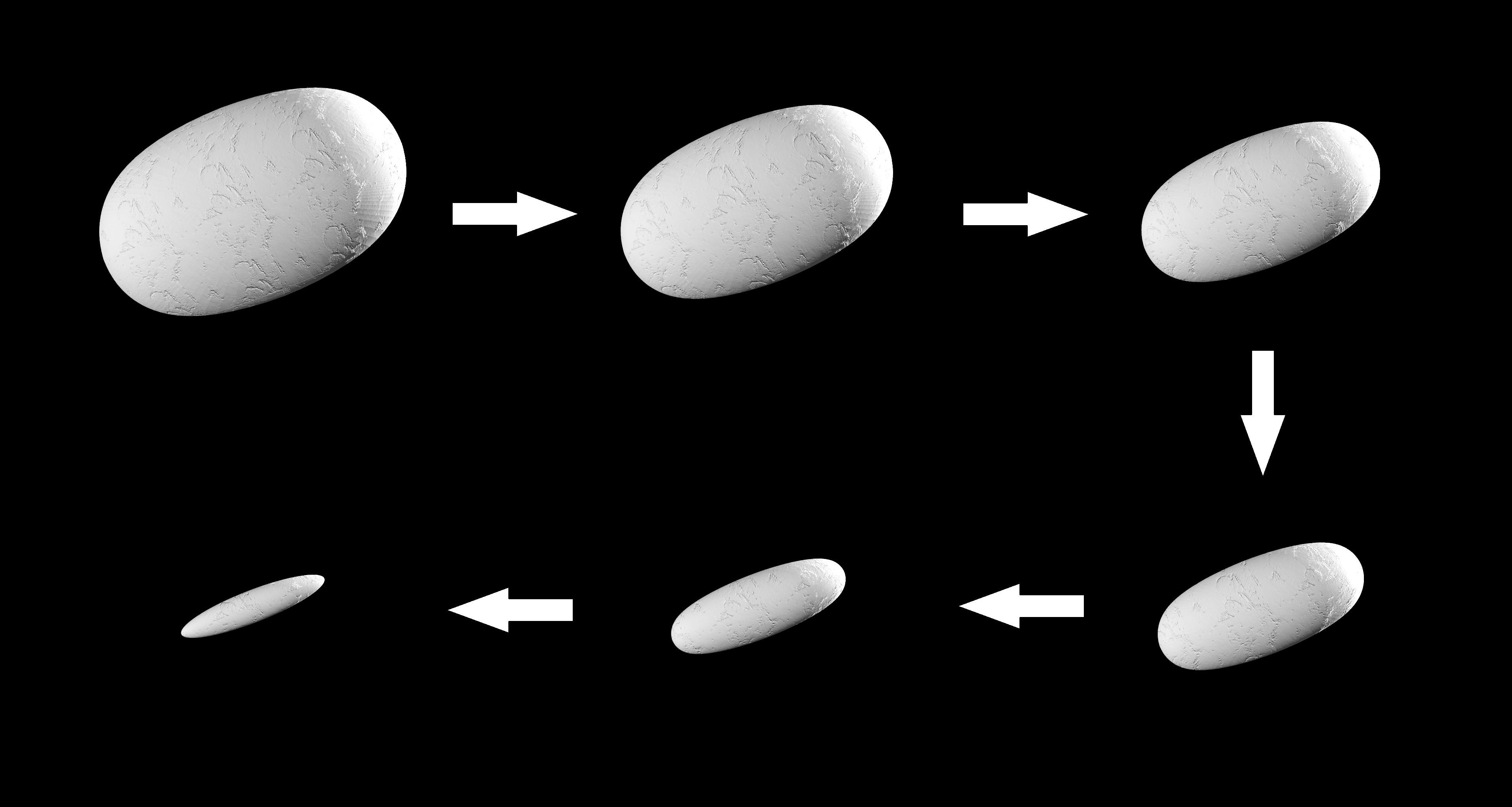}
		\caption{Transformation to an extremely elongated body by isotropic erosion.}
		\label{fig:Shape_change}
	\end{figure}
	
	Being the first confirmed interstellar object is not the only reason 'Oumuamua is unique. Its physical surface properties resemble those of cometary nuclei and D-type asteroids~\citep{Meech_etal2017,Fitzsimmons_etal2018,Jewitt_etal2017}. Although cometary activity could not be directly observed,~\citep{Meech_etal2017,Fitzsimmons_etal2018,Ye_etal2017}, a significant non-gravitational acceleration was discovered through orbital fitting.~\citep{Micheli_etal2018}. Exceeding two magnitudes, the light curve of 1I, too, is extraordinary suggesting a highly elongated shape. At the beginning the asteroid was modeled as a rotational ellipsoid with an axis ratio about 10:1~\citep{Meech_etal2017}. Later the ratio was slightly decreased to about 5:1~\citep{Jewitt_etal2017, Bannisteretal2017, Knight_etal2017} indicating that the shape resembled a ''cigar'' rather than a needle. Since such large axis ratios are basically unheard of among Solar System Objects,~\citep{McNeill_etal2017} it is not unreasonable to speculate that the asteroid may have acquired its peculiar form on its journey through interstellar space. 
	 {\citet{Hoang_etal2018} explained the elongated shape by spinup and  disruption of the asteroid caused by collisions with interstellar gas and consequent reassembly of binary fragments. In this work we consider another mechanism.}

	According to the latest research 'Oumuamua has a distant extrasolar origin and has been traveling in interstellar space for a long time~\citep{Mamajek2017, AlmeidaFernandes2018}. 
	Throughout this journey, the asteroid must have encountered gas molecules and interstellar dust grains at high relative speeds. Such high velocity collisions would over time result in significant surface ablation. Even though the densities of interstellar gas and dust are extremely low, high relative impact velocities in the Galaxy would lead to collisions that can significantly alter an asteroid's shape. 
	 {Here we propose two possibilities how the object could encounter the required mass of interstellar gas and dust: a) the interstellar asteroid is wandering the Galactic disk for a long time or b)  the interstellar asteroid passes through a relatively dense dust cloud.}
	
	The fact that 'Oumuamua is a 
	 {reliably determined}
	non-principle axis rotator~\citep{Fraser_etal2018,Drahus_etal2018,Belton_etal2018} 
	 {with 50~km/s dispersion velocity in the Galactic disk~\citep{Binney_Tremaine2008}}
	entails a near isotropic erosion of asteroidal surface material that gradually reduces the size of the asteroid. Since the asteroid is shrinking near uniformly on all sides, its axis ratio naturally increases. This process is schematically presented in Fig.~\ref{fig:Shape_change}. A $500\times300\times300$ meter sized asteroid with a 5:3 axis ratio exposed to isotropic erosion could, for instance, end up as a $250\times50\times50$ meter remnant having a new axis ratio of 5:1. 
	 {It should be noted that especially for case b) ''passing through a dust cloud'' a preferential direction of erosion could arise if deviation angle between the rotation axis and the principal axis is small.
		However, if the principle axis of rotation of a $500\times300\times300$ meter sized asteroid does not coincide with its long axis and the direction of erosion is not aligned with it the axis ratio can also increase. For instance, if the preferential direction of erosion is perpendicular to the precession axis, the principle axis of rotation coincides with the short axis and the precession angle is not high this asteroid exposed to isotropic erosion could end up as a $250\times250\times50$ meter object with again new axis ratio of 5:1.}
	
	
	In this work we estimate the amount of interstellar gas and dust necessary to erode 125 meters of asteroid material. How fast asteroid surface material is stripped off depends on an asteroid's physical properties as well as the relative speeds with respect to the gas and dust.

	\section{Erosion by gas molecules}

	In order to quantify how much material is ablated by incident gas molecules we used the open source software SRIM\footnote{\url{ http://www.srim.org/}}\citep{Ziegler_Manoyan1988} (The Stopping and Range of Ions in Matter). We considered molecules of Helium colliding with a Silicon target with density of $2 \ \mathrm{g/cm^3}$ at the speed of 100 km/s. In 1.4\% cases from 10000 simulations atoms of Silicon were knocked out. We used Helium in order to get the upper limit of the number of knocked out atoms. The mass of the Silicon surface ejected after collision, $m_{sur}$, is given by:

	\begin{equation}
	m_{sur}=0.014\frac{w_{Sil}}{w_{He}}m_g,
	\end{equation}
	
	\noindent where $m_g$ is the mass of Helium colliding with the target, $w_{Sil}$  and $w_{He}$ are the standard atomic weight of Silicon and Helium respectively. Here we considered the Silicon layer to be 125 m deep. Consequently the mass of Helium falling on $1\ \mathrm{cm^2}$ surface area is about 250 kilograms which is at least 8 orders of magnitude higher than the estimated mass of dust (see below).
	This fact justifies the assumption that the interstellar gas erosion is  negligible compare to collisions with interstellar dust grains even though the mass of interstellar dust is 100 times less than the gas mass in the Galaxy~\citep{Giese_1979}.

	\begin{figure}
		\includegraphics[width=\columnwidth]{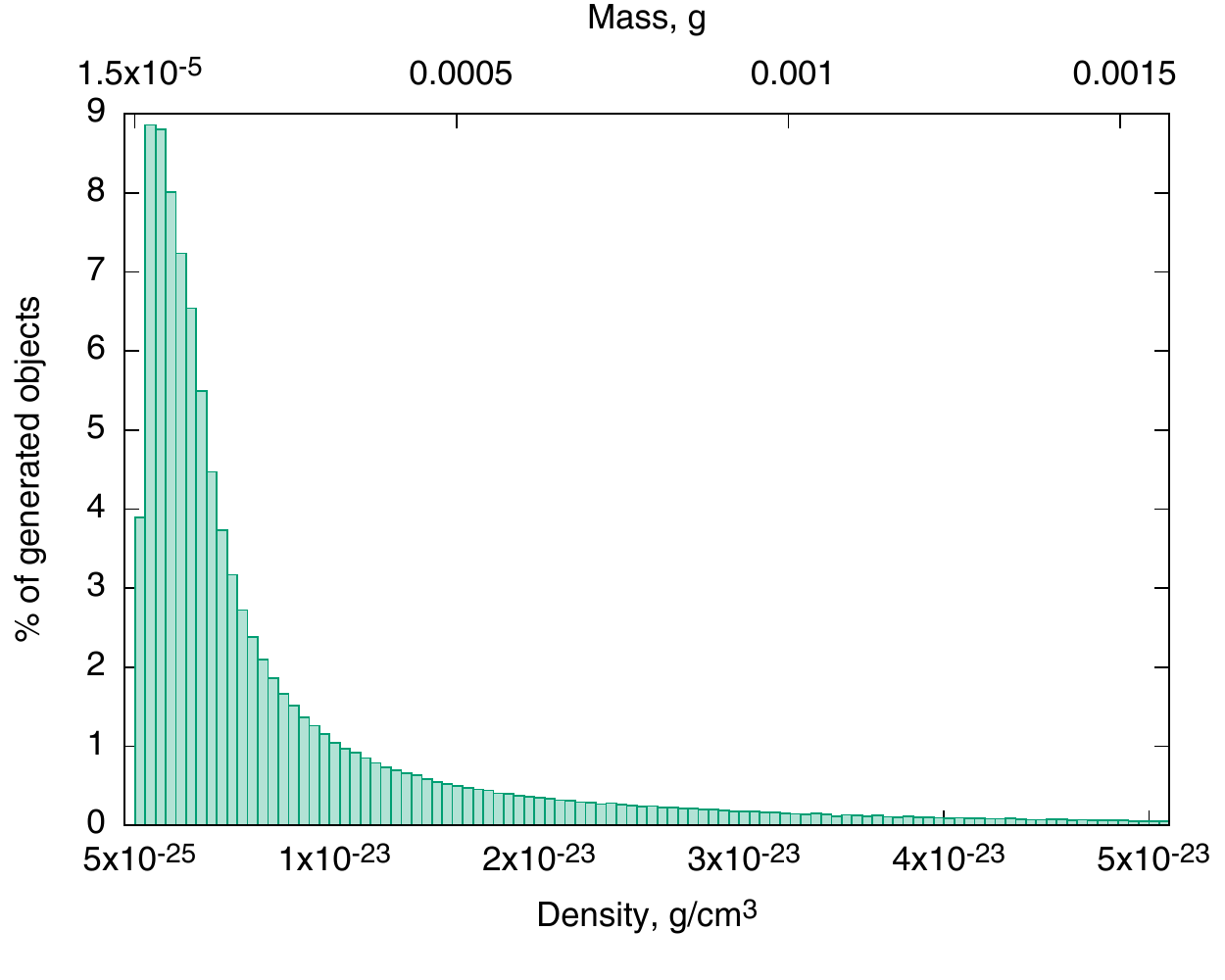}
		\caption{Histogram showing the mass of dust falling on  $1\ \mathrm{cm^2}$ of asteroid surface  (upper abscissa) and dust density of a 10 pc molecular cloud (lower abscissa) required to reduce the asteroid's size by 125 meters through isotropic and uniform ablation. The relative velocity, $v$, and the parameter $\mu$ were sampled from 20 to 200 km/s and from 0.51 to 0.55 respectively. }
		\label{fig:densities}
	\end{figure}

	\section{Erosion by dust grains}
	
	We estimated the required mass of dust to ablate 125 meters of the asteroid's surface by assessing the size of the impact craters after a dust grain collision. The radius of the crater $r$ reads~\citep{Housen_Holsapple2011}:
	
	\begin{equation}
	r=C_1 d \left( \frac{\rho_d}{\rho_a} \right)^{1/3} \left( \frac{v}{u} \right)^{\mu},
	\end{equation}
	
	\noindent where $C_1$  is a constant close to unity (equals 1.5 for rocks), $d$ is the radius of a dust grain, $\rho_d$  is the density of a dust grain,  $\rho_a$ is the asteroid's density, $v$  is the relative velocity of an asteroid with dust, $u$  is the escape velocity on the asteroid surface, $\mu$ depends on physical properties of the asteroid. Consequently, the mass of surface matter, $m_{sur}$, that will be ejected after collision is:
	
	\begin{equation}
	m_{sur}= \frac{C_1^3}{2} m_d  \left( \frac{v}{u} \right)^{3\mu},
	\end{equation}
	
	\noindent where $m_d$  is the mass of dust.  The change of the asteroid's size $dR$ is:
	
	\begin{equation}
	dR= \frac{C_1^3}{2\pi} \frac{m_d}{\rho_a} \left( \frac{v}{u} \right)^{3\mu}.
	\label{eq:size_change}
	\end{equation}
	
	\noindent For a spherical asteroid $u=R\sqrt{8\pi G \rho_a /3}$, where $R$ is the asteroid's radius. The $\pi$  factor arises from averaging over rotational period. Hence the equation~(\ref{eq:size_change}) is a differential equation the solution of which is: 
	
	\begin{equation}
	m_{dust} = \frac{1}{(3\mu + 1)B} \left( R^{3\mu + 1} - (R-\Delta R)^{3\mu + 1} \right),
	\end{equation}
	
	\noindent where
	\begin{equation}
	B = \frac{C_1^3}{2\pi\rho_a}v^{3\mu} \left( 8\pi G \rho_a /3 \right)^{-3\mu/2},
	\end{equation}
	\noindent $m_{dust}$  is the mass of dust that should fall on a unit area of the asteroid's surface to excavate $\Delta R$  layer. In this work we considered the asteroid's density as $\rho_a=2 \ \mathrm{g/cm^3}$, $C_1$ was set to be unity.

	\begin{figure}
		\includegraphics[width=\columnwidth]{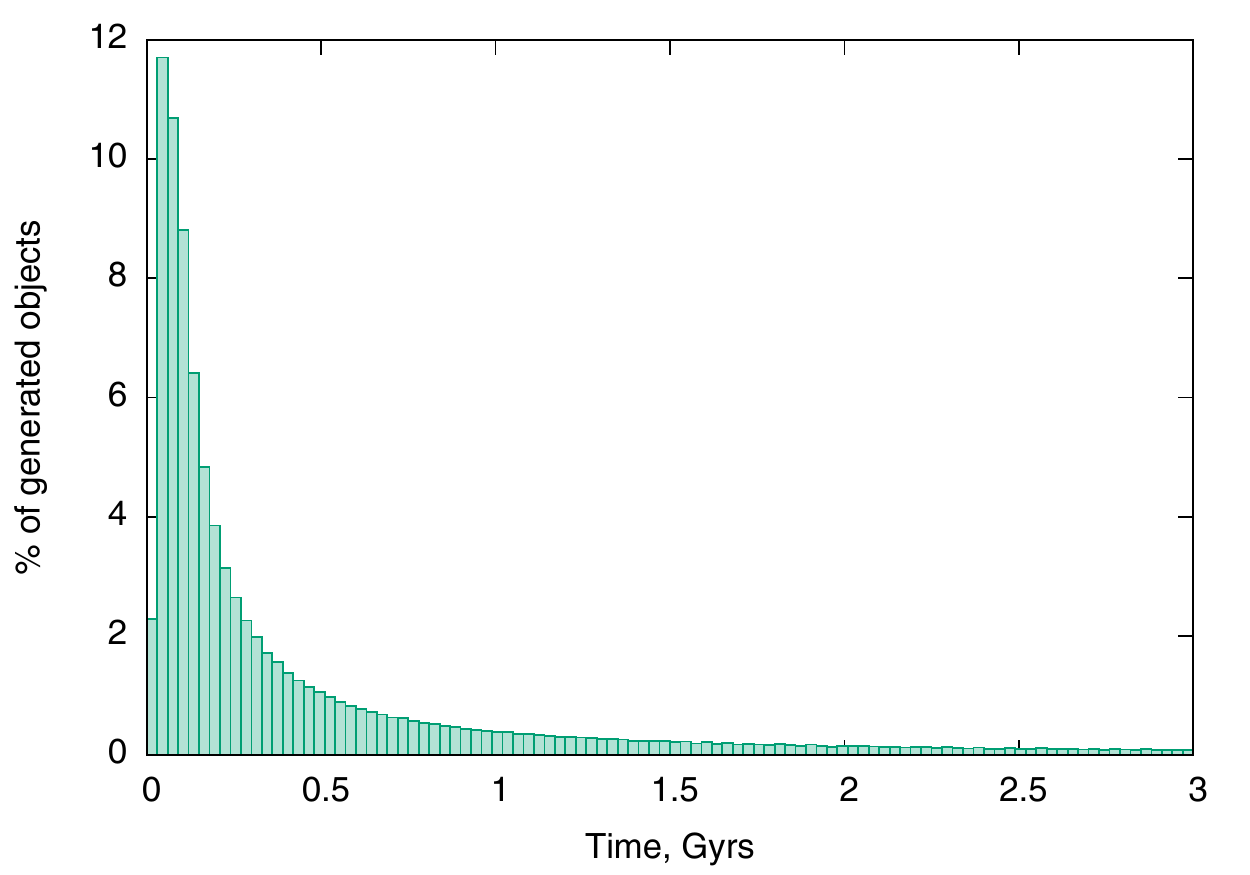}
		\caption{Galactic travel times required to shrink 'Oumuamua by 125~m equally from all sides. Asteroid parameter ranges are similar to those presented in Fig.~\ref{fig:densities}. }
		\label{fig:time}
	\end{figure}

	The most significant parameters are the relative speed, $v$  and the parameter $\mu$. The parameter $\mu$  depends on the physical properties of the asteroid. For rocks it equals approximately 0.55~\citep{Housen_Holsapple2011}, but could be a bit lower if the matter is more porous. Here we sample this parameter from 0.51 to 0.55. Since the 'Oumuamua's relative speed with respect to other stars ranges~\citep{Dybczynski_Krolikowska2018}  from 20 km/s to several hundreds of km/s we sample $v$ from 20 to 200 km/s. We estimate the mass of dust colliding with a surface area $1 \ \mathrm{cm^2}$ in size required to reduce the asteroid's size from all sides by 125 meters. We consider a rotational ellipsoid with 500 m major axis and 300 m minor axis as 'Oumuamua's initial shape. The results of the ablation process are shown in Fig.~\ref{fig:densities}. The required dust mass is between $10^{-5}$ and $10^{-3}$ grams depending on the relative speed and the asteroid's physical parameters.
	
	 {As it was said above} here we propose two possibilities how the object could encounter the required mass of interstellar dust: a) the interstellar asteroid is wandering the Galactic disk for a long time or b)  the interstellar asteroid passes through a relatively dense dust cloud. From the calculated dust mass we estimate the  density of a fictitious dust cloud 10pc in diameter (roughly 1/10th of the size of the Lagoon nebula). The results are presented in  Fig.~\ref{fig:densities}. The required dust grain density ranges from $5\cdot 10^{-25}$ to $5\cdot 10^{-23} \ \mathrm{g/cm^3}$, which is consistent with the observations~\citep{Ferriere2001}. The time it takes an asteroid to change its shape to an 1I like axis ratio ranges from millions to a couple of billion years (see Fig.~\ref{fig:time}), which is  reasonable taking into account that the planetary system 1I originated from is most likely not in the Solar neighborhood~\citep{Mamajek2017, AlmeidaFernandes2018}.
	 {It should be noted that the required time for slightly change of the asteroid shape (to erode 1~mm of the surface) are about 10 years in the ''b)'' case and from 3000 to 100000 years in the ''a)'' case. It is much less than the asteroid's rotation period, thus the averaging of the shape changes is reasonable.}
	
	Due to continuous collisions the asteroid's surface should be covered by a dust layer. The existence of radial non-gravitational acceleration while not having cometary coma on images~\citep{Meech_etal2017,Fitzsimmons_etal2018,Ye_etal2017} is explained by expulsion of coarse dust grains~\citep{Micheli_etal2018} which agrees with our study.

	\section{Lack of interstellar objects}

\begin{figure}
		\includegraphics[width=\columnwidth]{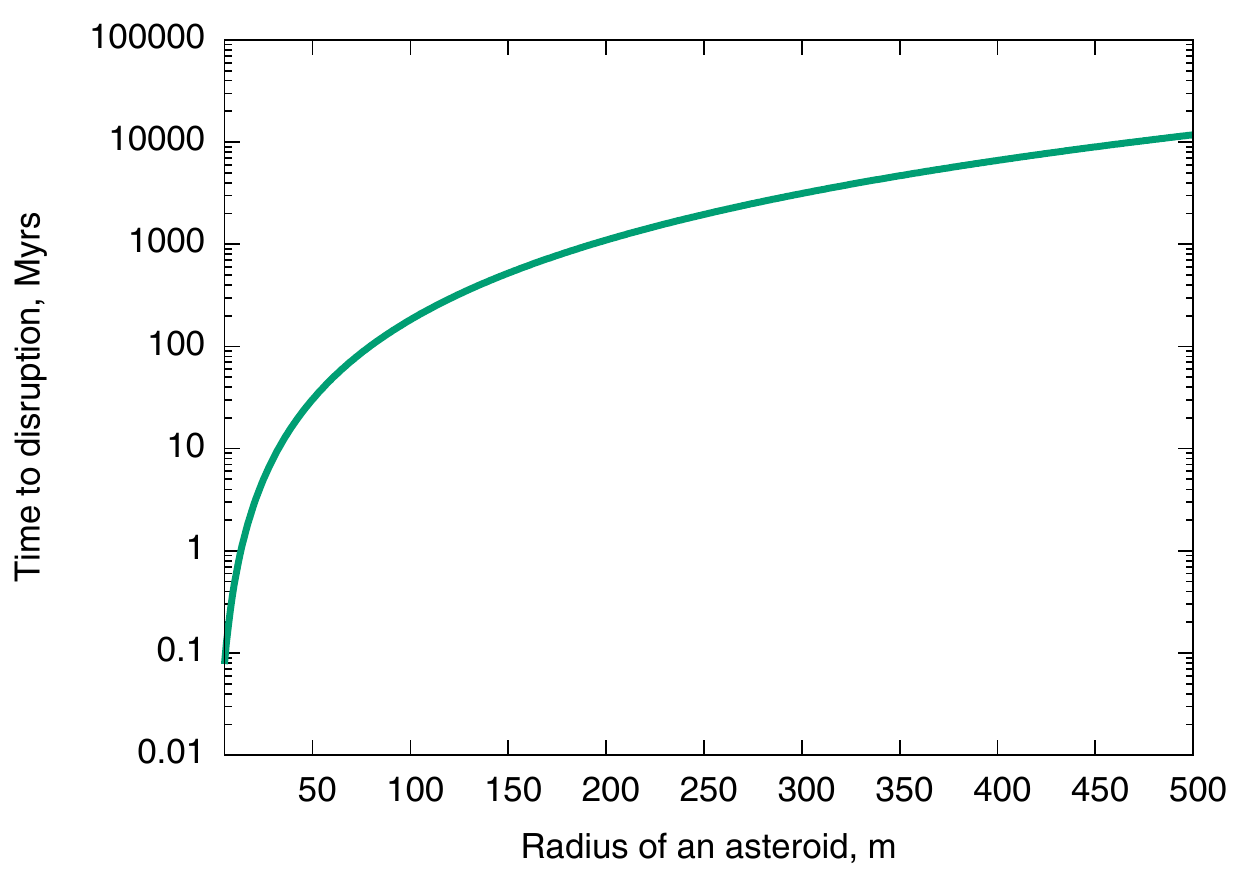}
		\caption{Disruption times for interstellar asteroids of different size ranges. Relative velocity $v=50 \ \mathrm{km/s}$ and $\mu = 0.53$. }
		\label{fig:lifetime}
	\end{figure}

	Studies on the formation and evolution of the Solar system suggest that 99\% of the original planetasimals in planetary systems are ejected by orbital migration of giant planets and gas~\citep{Charnoz_Morbidelli2003,Bottke_etal2005}. As a consequence, interstellar space should be filled with interstellar objects. The observed number of such objects by solar system surveys as Pan-STARRS1, the Mt. Lemmon Survey, and the Catalina Sky Survey should be substantially greater~\citep{Engelhardt_etal2017}. 
	In fact, 'Oumuamua was the first 'interstellar vagabond' discovered by solar system surveys despite decades of near-continuous observation. To shed some light on the solution to this conundrum we estimated the averaged lifetime of an asteroid (the time required to completely destroy the asteroid by dust grains impacts in Galactic disc) as a function of the asteroid's size. The results are shown in Fig.~\ref{fig:lifetime}. As one can see an asteroid, with an initial diameter of 100 meters, can wander the Galactic plane for no longer than 30 million years before it is 'sandblasted' into tiny fragments.
	This fact may explain why interstellar objects are relatively rare and it should be taken into account in future assessment of the number of observed interstellar objects. This mechanism also reveals the source for interstellar dust, the origin of which is not completely understood yet~\citep{Ferriere2001}.

	Such short lifetimes, compare to the Solar system age, for small interstellar objects should not be considered as lifetimes of main belt asteroids. The size of interstellar dust grains ranges from $0.1 \ \mathrm{\mu m}$ to $0.5 \ \mathrm{\mu m}$~\citep{Draine2009} and hence they are strongly affected by Solar pressure. In their paper \citet{Mann_Kimura2000} indicate the gap in mass distribution of interstellar dust in the Solar system at distances $r < 3 \ \mathrm{au}$, which is caused by radiation pressure repulsion. Because of that the  asteroid's surface erosion  by dust grains, described above, affects main belt asteroids significantly less. Nevertheless the objects at large distances from the Sun (such as long periodic comets and Trans-Neptunian objects) should be exposed by this erosion fully. Consequently we predict the extinction of small bodies of this kind, which is partially confirmed by shortage of small long periodic comets. Also 'Oumuamua's surface properties are alike to D-type asteroids, which are suggested to be originated in the Kuiper belt~\citep{McKinnon2008} and suffered from dust erosion.

	\section{Summary}

	The only interstellar object discovered so far, 'Oumuamua turned out to be extremely elongated experiencing a relatively large non-gravitational acceleration without showing signs of significant cometary activity.
	Our simulations show that asteroid collisions with interstellar dust grains can cause a near isotropic erosion of bodies in non-priniciple axis rotation states that will reduce the asteroid's size significantly over time. If a body was slightly elongated when it left its parent planetary system the isotropic erosion increases the axis ratio and, thus,  makes the body more elongated. 'Oumuamua could have evolved into its current shape during 20 Myrs -- 2 Gyrs of interstellar travel in the Galactic disk. Alternatively the asteroid could have passed through an interstellar dust cloud which would have had a similar ablation effect. 
	Collisions with interstellar dust can be a reason for the existence of course dust on the surface of 'Oumuamua, which is believed to cause significant non-gravitational acceleration. The shortage of observed interstellar objects is explained by relatively short time (tens of millions of years) a 100 m size asteroid can travel in the Galactic disk without being destroyed.

	\section*{Acknowledgements}
	We are very grateful to Dr. Siegfried Eggl for helpful comments and the first review of the paper.
	This work was supported by a grant of Russian Science Foundation \#16-12-00071.
	
	
	
	


	
	
	
	
	

	\bsp	
	\label{lastpage}
\end{document}